\documentclass[a4paper]{panl}
\usepackage{cite}
\usepackage{wrapfig}
\usepackage{graphicx}
\usepackage{amssymb}
\usepackage{amsfonts}
\usepackage{amsmath}
\usepackage{longtable}
\usepackage{rotating}
\usepackage{lscape}
\usepackage{epsfig}
\usepackage{multirow}

\usepackage{lineno} 

\originalTeX
\begin{document}
\issuearea{Physics of Elementary Particles and Atomic Nuclei. Theory}


\title{Physics and astrophysics of ultra-high energy cosmic rays: recent results from the Pierre Auger Observatory}
\maketitle
\authors{Jo\~ao R. T. de Mello Neto$^{a,b,}$\footnote{E-mail: jtmn@if.ufrj.br}, 
{ for the Pierre Auger Collaboration} $^{b,}$\footnote{ Full author list: http://www.auger.org/archive/authors\_2020\_10.html}}
\from{$^{a}$\, Instituto de F\'\i sica, Universidade Federal do Rio de Janeiro, Ilha do Fund\~ao\\
Rio de Janeiro, RJ  Brazil}
\vspace{-3mm}
\from{$^{b}$\, Observatorio Pierre Auger, Av. San Mart\'\i n Norte 304, 5613 Malarg\"ue, Argentina}

\begin{abstract}
\vspace{0.2cm}

Ultra-high-energy cosmic rays (UHECRs) are the highest energy messengers in the universe, with energies up to $10^{20}$ eV.  Studies of astrophysical particles (nuclei, electrons, neutrinos and photons) at their highest observed energies have implications for fundamental physics as well as astrophysics. The primary particles interact in the atmosphere (or in the Earth) and generate  extensive air showers. Analysis of those showers enables one not only to estimate the energy, direction and most probable mass of the primary cosmic particles, but also to obtain information about the properties of their hadronic interactions at energies more than one order of magnitude above that accessible with the current highest energy human-made accelerator.  
The Pierre Auger Observatory, located in the province of Mendoza, Argentina, is the largest cosmic ray experiment ever built. The observatory was designed as a hybrid detector covering an area of 3000 km$^2$ and  has been taking data for almost twenty years. 
In this paper, a selection of the latest results is presented:  the cosmic ray energy spectrum, studies of hadronic physics, searches for a directional anisotropy and studies of mass composition (including the photon and neutrino searches).  Finally, the current upgrade (``AugerPrime'') of the observatory, which is  mostly aimed at improving the sensitivity to the particle type and mass of ultra-high energy cosmic rays, is described.


\end{abstract}
\vspace*{6pt}


\label{sec:intro}
\section*{Introduction}
Cosmic rays are particles arriving from outer space incident on the Earth's atmosphere (primaries), e.g. protons, heavier nuclei, photons and neutrinos plus the particles generated when they interact in the atmosphere (secondaries).   Studies of astrophysical particles   at their highest observed energies have implications for fundamental physics as well as astrophysics. The existence of Ultra High Energy Cosmic Rays   (UHECR - here standing for cosmic rays with energy $E \geq 0.1$ EeV, 1~EeV $\equiv 10^{18}$~eV)  has been known since the 1960s\cite{PhysRevLett.6.485,Linsley:1963km} but  their  nature and origin remain mysterious.     
For energies up to $10^{15}$~eV, cosmic rays probably have a galactic origin and the most likely mechanism could be  shock acceleration in supernova remnants. At the highest energies, the most probable sources of  UHECRs  are extragalactic:   jets of active galactic nuclei (AGN), radio lobes, gamma-ray bursts and colliding galaxies, among others \cite{Kotera2011}.
The UHECRs are the subject of extensive studies, including measurements of the energy spectrum, the nature of the primaries, and identification of potential sources. With energies that can reach about two orders of magnitude higher than the ones possible in the LHC, the study of the UHECRs probes the hadronic interactions in otherwise inaccessible regions of the phase space. 

The Pierre Auger Observatory is composed of several systems.  It is a hybrid detector that combines both surface and fluorescence detectors at the same site.  The Surface Detector Array (SD) consists of 1660 10\,m$^2$\,$\times$\,1.2\,m   water-Cherenkov detectors (WCD) deployed over 3000
km$^2$ on a 1500 m triangular grid (SD1500).  A smaller array with 61 detectors on a  750 m grid was added to the SD with the purpose of measuring showers of  lower energy (SD750). 
The SD is overlooked by a fluorescence detector (FD) composed of twenty-four fluorescence telescopes, grouped in units of six wide-angle telescopes   at four  buildings on its periphery.  In addition, showers of lower energy are measured   in an  additional building  with three high-elevation telescopes (HEAT).  The surface detector   stations
sample   the electrons, photons and muons   in the shower front at ground level. The fluorescence telescopes  can record   ultraviolet  light emitted as the shower crosses the atmosphere, allowing  one  to observe  the longitudinal development of the air shower.
The fluorescence detector operates only on clear, moonless nights, so its duty cycle is about 13\%. On the other hand, the  surface detector array has a duty cycle close to 100\%. 
The Auger Engineering Radio Array (AERA) complements the measurements of low energy showers by detecting their radio emission, using more than 150 radio antennas. Long term  monitoring of the detectors and real-time monitoring of the atmosphere are performed with different atmospheric monitoring devices \cite{Harvey:2019gH, Choi:201982}. 
A  detailed description of the observatory and  of the  methods used to determine the energy and arrival direction from the data   has been published \cite{ThePierreAuger:2015rma}.

The collaboration is composed of about 400 scientists from 17 countries and a very dedicated staff at the  observatory.  The observatory is well connected with the local population, and the outreach activities are a key factor in the success of the project \cite{Timmermans:2017x7,Garcia:2019y3}. 

\label{sec:spectrum}
\section*{The cosmic ray energy spectrum}

The all-particle energy spectrum is the most outstanding observable in cosmic ray physics, since it contains information in a combined way about the sources and about the galactic and/or intergalactic media in which the cosmic rays propagate.   
The energy spectrum of cosmic rays above $2.5 \times 10^{18}$ eV was measured based on 215\,030 events recorded by the SD. The total signal at a core distance of 1000 m, $S(1000)$, is the SD energy estimator.  The observatory is a  hybrid system, and so its  energy scale can be set with the FD measurements that provide an almost calorimetric estimate of the shower energy.  This procedure allows measuring the energy spectrum with an energy estimation which is largely independent of air shower simulations and of assumptions on hadronic interaction models.
The spectrum multiplied by $E^3$ is shown in Fig.~\ref{f1} superimposed by a sequence of four power-laws fitted to the data. The values of the spectral index $\gamma$  are shown here confirming  the flattening of the flux near $5\times10^{18}$ eV, the so-called "ankle", and the abrupt suppression at around $5\times10^{19}$ eV. The flux also shows a softening  at about $1.3 \times 10^{19}$ eV, calling for a two-step suppression. This feature was never observed previously.   Another new observation is that the spectrum shows no declination dependence \cite{Aab:2020gxe}.  

To study the astrophysical implications of the features of the energy spectrum,  a simple scenario was used in which a few nuclear componentes are injected at the sources with a power-law spectrum and with the maximal energy of the sources modelled with an exponential cutoff.  The sources are  assumed to be stationary and uniform in a comoving volume. 
Using the data from the energy spectrum and  mass composition  \cite{Aab:2016htd,Yushkov:2019J8} we show in Fig.~\ref{f2} the best reproduction of the data by simultaneously fitting the energy spectrum above $5 \times 10^{18}$~eV  and the distribution of the depths of the shower maximum ($X_{\textrm{max}}$), which is a mass-sensitive variable measured  using the FD.   EPOS LHC \cite{PhysRevC.92.034906}  was used as a model of hadronic interactions in the interpretation of the results. In this scenario, the  intermediate-mass nuclei accelerated to $\approx 5 Z \times 10^{18}$  eV and escaping from the source environments with a very hard spectrum dominate the nuclear abundance at the sources.  The steepening observed above $5 \times 10^{19}$ eV is caused by the combination of the maximum energy of acceleration of the heaviest nuclei at the sources and the energy losses during the propagation (the GZK effect).  The steepening at $\approx 10^{19}$ eV is due to the cutoff of the helium spectrum with CNO contribution shaped by the photodisintegration effect \cite{Aab:2020rhr}.

\begin{figure}[h]
\begin{minipage}{16pc}
\includegraphics[width=16pc]{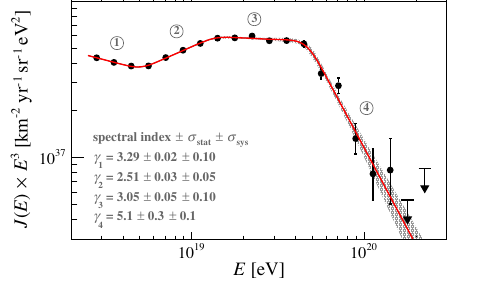}
\caption{Energy spectrum scaled by $E^3$ fitted with a sequence of four power laws. The red line is the fit. The numbers from 1 to 4 enclosed in the circles identify the energy intervals where the spectrum is described by the power law with the respective index \cite{Aab:2020rhr}.}
 \labelf{f1} 
\end{minipage}\hspace{1pc}%
\begin{minipage}{16pc}
\includegraphics[height=10pc,width=16pc]{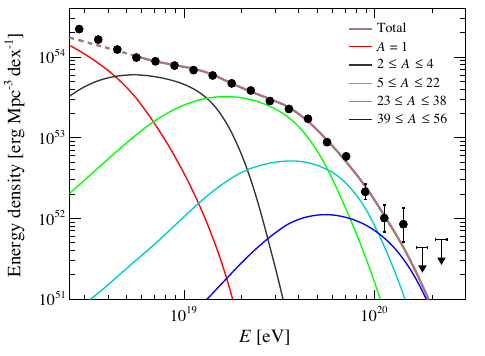}
\caption{Energy density obtained with the best fit parameters of the benchmark scenario described in the text. The dashed curve shows the energy range that is not used in the fit and where an additional component is needed for  describing the spectrum \cite{Aab:2020rhr}.}
 \labelf{f2} 
\end{minipage} 
\end{figure}

\label{sec:comp}
\section*{Mass composition measurements}
As we mentioned above, $X_{\textrm{max}}$, the depth of the shower maximum, is a mass-sensitive variable because  showers from heavier nuclei develop higher in the atmosphere and their profiles fluctuate less, while showers from lighter nuclei develop deeper in the atmosphere and their profiles fluctuate more. 
The $X_{\textrm {max}}$ of proton showers are on average about 100~g/cm$^2$ deeper in the atmosphere than the $X_{\textrm {max}}$ of iron showers. In a similar way, the fluctuation of the values of $X_{\textrm {max}}$ around the mean depth of the shower maximum, $\textrm{RMS}(X_{\textrm {max}})$, provides another sensitive observable thought it depends on both the average and the spread of the mass distribution. Iron showers fluctuate about 40 g/cm$^2$ less than proton showers.
 Measurements of   $X_{\textrm{max}}$ are performed at energies above $10^{17.8}$ eV with the standard FD telescopes. The HEAT telescopes make it possible to measure showers with energies down to $10^{17.2}$ eV.  The $X_{\textrm{max}}$ moments are free from detector effects and can be compared to predictions from MC simulations in a direct way.  The resolution of the standard FD  is $\sim$\,25\,g/cm$^2$ at $10^{17.8}$ eV and  at high energies it is $\sim$\,15\,g/cm$^2$.  For most of the energy range the systematic uncertainties are below 10\,g/cm$^2$ \cite{Aab:2014kda}.
In Fig.~\ref{f3} the measurements of $\langle X_{\textrm {max}} \rangle$ as a function of energy are shown, and in Fig.~\ref{f4}  $\sigma(X_{\textrm {max}}) $ in function of energy is also shown.  The mean mass of the UHECRs shown in Fig.~\ref{f3} decreases as function of energy until $E_0=10^{18.32 \pm 0.03}$ eV and increases for higher energies. The smaller values of $\sigma(X_{\textrm {max}}) $  for energies above $E_0$ shown in Fig.~\ref{f4} are as well in agreement with the MC predictions for  $\sigma(X_{\textrm {max}}) $ of heavier nuclei.
The results agree with previous measurements of the same observables \cite{Abraham:2010yv, Aab:2014kda, Bellido:2017Li}.

\begin{figure}[h]
\begin{minipage}{16pc}
\includegraphics[width=16pc]{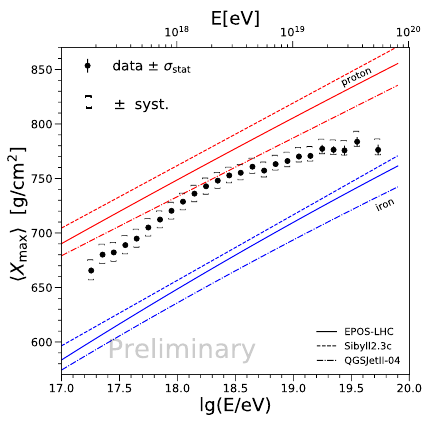}
\caption{ $\langle X_{\textrm {max}} \rangle$ as a function of energy   compared to air-shower simulations for proton and iron primaries using the hadronic models EPOS-LHC, Sibyll 2.3c and QGSJetII-04  \cite{Yushkov:2019J8}.}
 \labelf{f3} 
\end{minipage}\hspace{1pc}%
\begin{minipage}{16pc}
\includegraphics[width=16pc]{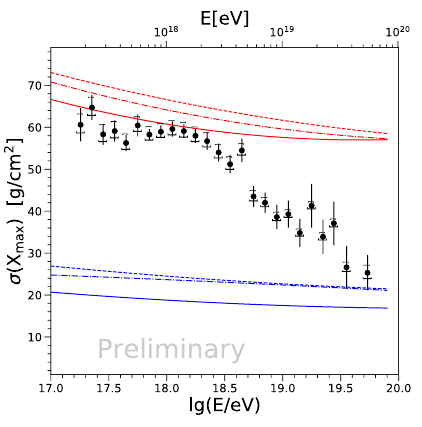}
\caption{  $\sigma(X_{\textrm {max}}) $ as a function of energy   compared to air-shower simulations for  proton and iron primaries using the hadronic models EPOS-LHC, Sibyll 2.3c and QGSJetII-04   \cite{Yushkov:2019J8}.}
 \labelf{f4} 
\end{minipage} 
\end{figure}

The correlation between  $X_{\textrm {max}}$ (measured by the FD)  and the signal in the  WCD  is another measurement relevant for mass composition\footnote{The measurement must be performed with the usage of two independent detector systems to avoid correlated detector systematics.}.  With this correlation the spread of the masses in the primary UHECRs can be estimated.  In \cite{YOUNK2012807} the correlation between $X_{\textrm {max}}$ and the number of muons $N_{\mu}$ in air showers was proposed as an observable to determine whether the mass composition is pure or mixed. Correlations close to or larger than zero are found in simulations for pure cosmic-ray mass  composition.  In contrast, a negative correlation is obtained in mixed mass composition.  The data used for this measurement has $10^{18.5}  \leq E/\textrm{eV} \leq 10^{19.0} $. The surface array of WCD provide a significant sensitivity to muons: for zenith angles between 20 and 60 degrees, muons contribute about 40\% to 90\% of $S(1000)$.  So $S(1000)$ is used in the place of $N_{\mu}$.  To avoid a decorrelation due to the spreads of energies and zenith angles, the observables $X_{\textrm {max}}$ and $S(1000)$ are scaled to a reference energy of 10 EeV.   In addition $S(1000)$ is scaled to a zenith angle of $38^{\circ}$. The rescaled variables are denoted as $X^{\ast}_{\textrm {max}}$ and $S^{\ast}_{38}$.  The correlation for a pure proton  sample is $r_G = 0.04$ and for a pure iron sample is $r_G=0.12$, with both Monte Carlo samples produced with the EPOS-LHC model.  But for the data the correlation is $r_G = -0.069 \pm 0.017$.  The negative correlation found in data cannot be reproduced using any pure composition.  In Fig.~\ref{f5} the dependence of the simulated correlation $r_G$ on the spread of $\sigma(\ln{A})$ is shown for EPOS-LHC. The correlation found in data is compared to the values in simulated mixtures with all possible combinations of relative fractions of (p, He, O, Fe) nuclei changing with a step of 0.1.  The spread of the primary masses $\sigma(\ln{A})$ can be estimated from Fig. ~\ref{f5} to be $0.85  \lesssim \sigma(\ln{A}) \lesssim 1.6 $.  
The comparison of the energy dependence of $r_G$ in data to the predictions for proton, iron and extreme mix p/Fe = 1/1 for EPOS-LHC interaction model   is shown in Fig.~\ref{f6}.  For higher energies, the correlation in data becomes consistent with the compositions with smaller mixings.
 This work was reported by the Pierre Auger Observatory in \cite{Aab:2016htd} and updated in \cite{Yushkov:2019J8}.

\begin{figure}[h]
\begin{minipage}{16pc}
\includegraphics[width=16pc]{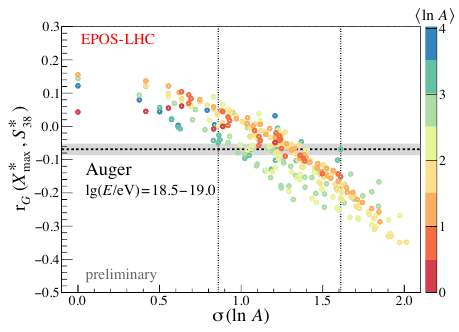}
\caption{ Dependence of the correlation coefficients $r_G$ on $ \sigma(\ln{A})$ for EPOS-LHC. Each simulated point corresponds to a mixture with different fractions of (p, He, O, Fe), the relative fractions changing in steps of 0.1.  Four points in the upper left show the pure compositions at $ \sigma(\ln{A}) = 0 $   \cite{Yushkov:2019J8}.}
 \labelf{f5} 
\end{minipage}\hspace{1pc}%
\begin{minipage}{16pc}
\includegraphics[width=16pc]{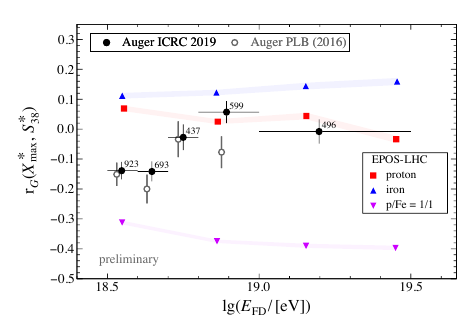}
\caption{The correlation coeficients $r_G$ for data (full circles) in five bins.  Results from our previous publication \cite{Aab:2016htd} are shown with open circles.  The two datasets are statistically compatible \break ($p$-value = 0.25).  Predictions are given for EPOS-LHC
 \cite{Yushkov:2019J8}. }
 \labelf{f6} 
\end{minipage}
\end{figure}

\pagebreak
\label{sec:hadronic}
\section*{Hadronic physics} 
Analysis of UHECRs enables one to obtain information about the properties of their hadronic interactions at an energy more than one order of magnitude above that accessible with the current highest energy human-made accelerator.

\begin{figure}[h]
\begin{minipage}{16pc}
\includegraphics[height=10pc,width=16pc]{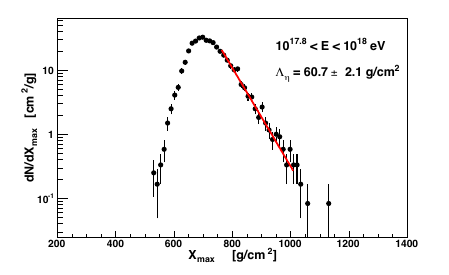}
\caption{ The result of the unbinned log-likelihood fit to derive $\Lambda_{\eta}$ is shown in the range of the tail of the $X_{\textrm{max}}$ distribution fit for the first bin in energy ranging from $10^{17.8}$ to $10^{18}$ eV  \cite{Ulrich:2016oO}.}
 \labelf{f7} 
\end{minipage}\hspace{1pc}%
\begin{minipage}{16pc}
\includegraphics[height=10pc,width=16pc]{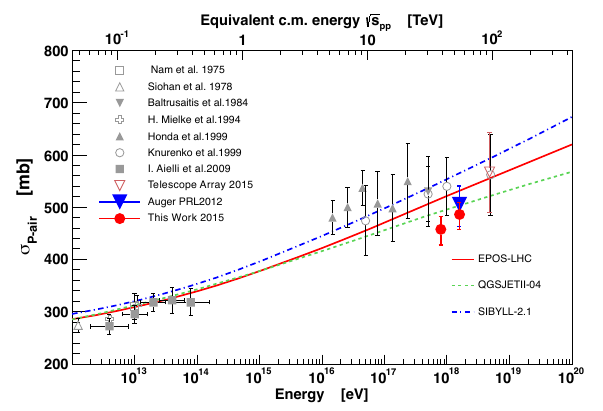}
\caption{  Resulting $\sigma_{p-air}$ compared to other measurements and model predictions. The blue point was reported in \cite{Collaboration:2012wt} and the two red points are from the updated analysis in  \cite{Ulrich:2016oO}. }
 \labelf{f8} 
\end{minipage}
\end{figure}

An update of the analysis of the proton-air cross-section based on the shape of the distribution of $X_{\textrm{max}}$ published  in \cite{Collaboration:2012wt} was presented in \cite{Ulrich:2016oO}.  The analysis is based on the fact that the tail of the $X_{\textrm{max}}$ distribution is sensitive to $\sigma_{\textrm{p-air}}$.  The 20\% most deeply penetrating showers were selected for the analysis to reduce the impact of primary cosmic ray nuclei heavier than protons. 

The cross section is  related to the exponential distribution of the depth of the first interaction $X_1$ which cannot be measured.  But the strong correlation between $X_1$ and $X_{\textrm{max}}$ makes the distribution of the latter sensitive to the proton-air cross-section and the tail of the distribution maximizes the proton content, since it is the most penetrating nucleus.   The slope  ( $\Lambda_{\eta}$) obtained from a fit to the exponential tail of the  $X_{\textrm{max}}$ distribution (see Fig.~\ref{f7}) can be used as an estimator for $\sigma_{p-air}$ through Monte Carlo simulations: the cross-section is rescaled consistently to reproduce the value of the measurement. The lack of detailed knowledge of the mass composition at these energies turns out to be the main difficulty for this analysis since one cannot exclude contamination by photons and helium primaries, for instance. This translates into the main contribution to the systematic uncertainty of this measurement. 

The available data sample is divided into two energy intervals, one with 18090 events ranging from $10^{17.8}$ to $10^{18}$ eV and the other  21270 events from $10^{18}$ to $10^{18.5}$  eV   and the measured cross-sections are $457.5 \pm 17.8 ({\rm stat})  ^{+19}_{-25} ({\rm sys})$ mb and $485.8 \pm 15.8 ({\rm stat})  ^{+19}_{-25} ({\rm sys})$ mb respectively. 
While the composition of primary cosmic rays in the above energy ranges is compatible with being dominated by protons, a contamination with helium cannot be excluded.  The quoted systematic uncertainties take into account, among many other effects, an impact of 25\% helium in the data sample. 
Fig.~\ref{f8} displays the $\sigma_p$-air measurement compared to previous data and model predictions.  The data are consistent with a rising cross section with energy, however, the statistical precision is not yet sufficient to make a statement on the functional form \cite{Ulrich:2016oO}.

\begin{figure}[h]
\begin{minipage}{16pc}
\includegraphics[height=12pc,width=15pc]{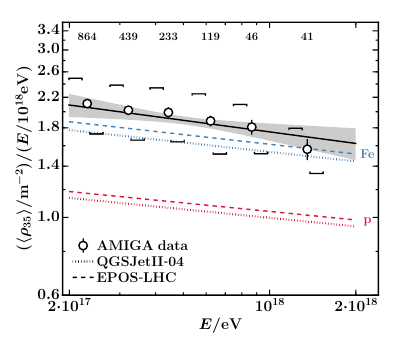}
\caption{ Energy-normalized densities as a function of $E$ compared to expectations. Error bars denote the statistical uncertainties. Systematic uncertainties are indicated by square brackets.  The   fit is represented by the black solid line, the  shaded band shows the statistical uncertainties \cite{Sanchez:2019KN}.}
 \labelf{f9} 
\end{minipage}\hspace{1pc}%
\begin{minipage}{16pc}
\includegraphics[height=12pc,width=16pc]{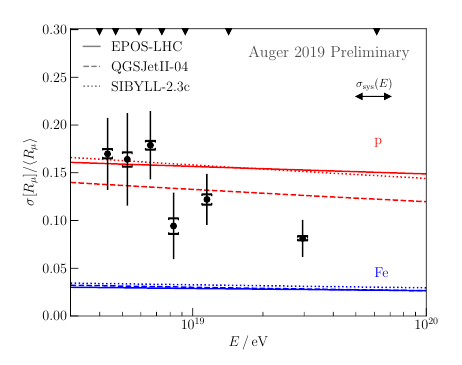}
\caption{  Shower to shower fluctuations. The statistical uncertainty (error bars) is dominant.  The systematic effects are shown by the square brackets. The energy ranges for which the fluctuations are evaluated are marked by the black triangles at the top of the figure \cite{Riehn:20194u}. }
 \labelf{f10} 
\end{minipage}
\end{figure}

The measurement of the muonic component at the ground is very sensitive to the characteristics of the hadronic interactions along many steps of the cascade,  such as the fraction of the electromagnetic component of the shower with respect to the total signal and the multiplicity of the secondaries \cite{Cazon:2018gww}.

The Pierre Auger Collaboration has been showing consistently that the number of muons  in the models is smaller than what is measured in data \cite{Aab:2016hkv}. More recently we showed that this inconsistency between data and models is also present at lower energies, by directly measuring the muon content of air showers with an engineering array of underground muon detectors (UMD) deployed in the SD750 area.  

In figure Fig.~\ref{f9}  we see the evolution with energy of the muon content in data compared to simulations with proton and iron primaries. To soften the strong energy dependence, the muon densities have been normalised by  the energy.  The slopes in the curves for the hadronic models are 0.91 for iron and 0.92 for proton, slightly steeper than the slope for data $ b = 0.89 \pm 0.04 (\textrm{stat}) \pm 0.04 (\textrm{sys}) $.    Additionally, simulations fail to reproduce the observed muon densities which are between 8\%  for the EPOS-LHC model and 14\% for the QGSJetII-04 larger than those obtained for iron showers at an energy of  $10^{18}$ eV \cite{Sanchez:2019KN}.                                                                                                                                                       

Another observable of UHECRs that can constrain exotic models to explain the muon excess is the shower-to-shower fluctuations in the number of muons. This measurement is performed with inclined air showers since the electromagnetic cascade is heavily absorbed in the atmosphere and the signals at the ground are dominated by muons.   A fit of the normalization factor of a reference model for the muon density at the ground to the observed distribution of signals in the SD array provides the number of muons at the ground relative to the average of the total number of muons  in a shower with primary energy of $10^{19}$ eV.   In Fig.~\ref{f10} we show the results of the relative fluctuations in the number of muons.  The observed fluctuations fall in the range of the predictions from air shower simulations with current hadronic interaction models.  The measured fluctuations decrease with primary energy:  fitting $ p_0 + p_1 \log_{10}{(E/\textrm{eV})}$ to the fluctuations we find $p_1 = -0.11 \pm 0.04$.  Thus, the results suggest that models are describing reasonably well the distribution of energy going into the electromagnetic component after the first interaction, while the discrepancy in the overall muon number should be explained by resorting to changes of different hadronic interaction characteristics along all the shower stages                  
\cite{Riehn:20194u}.

\label{sec:anisotropy}
\section*{Arrival direction anisotropy}

An observable that sheds light on the nature and origin of UHECRs is the distribution of their arrival directions over the sky.  Their arrival directions are basically free from systematic errors, in contrast with energies or primary mass, but do not exactly correspond to the position of the sources because of deflections of cosmic rays by intergalactic and galactic magnetic fields. Although the sources of UHECRs are yet to be discovered, their large-scale distribution is expected to follow the local distribution of matter in the universe at some level.  Dipoles, quadrupoles and higher order multipoles of the distribution in the sky could be present due to diffusive propagation of UHECRs, excesses in the supergalactic plane, and other possible features of the source distributions.
\begin{figure}[h]
\begin{minipage}{18pc}
\includegraphics[width=18pc]{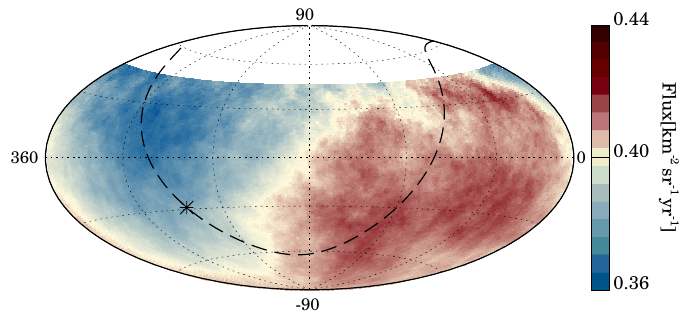}
\caption{The CR flux above 8 EeV, averaged on top-hat windows of $45^{\circ}$ radius in equatorial coordinates. The Galactic plane is indicated by a dashed line  and the Galactic center by a star  \cite{Roulet:2019tW}.}
 \labelf{f11} 
\end{minipage}\hspace{1pc}%
\begin{minipage}{14pc}
\includegraphics[height=8.5pc,width=14pc]{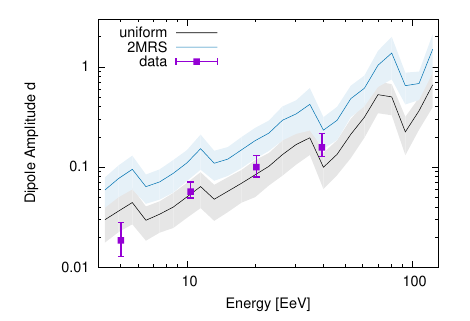}
\caption{  Energy dependence of the dipolar amplitude measured above 4 EeV. Also shown are the predictions from scenarios with extragalactic sources \cite{Roulet:2019tW,Harari2015}. }
 \labelf{f12} 
\end{minipage}
\end{figure}

In \cite{Aab:2017tyv}  a dipolar anisotropy was reported in the arrival directions of UHECRs above $ 8 \times 10^{18}$ eV and it was updated in \cite{Roulet:2019tW}.  The data cover more than three decades in energy and comprises about 15 years of measurements.  Above the full trigger efficiency of the array, a weighted Fourier analysis in right-ascension and azimuth and below 2~EeV the East-West method was used down to 0.3~EeV.  As  can be seen in Fig.~\ref{f11}, for $E > 8$~EeV  a clear dipolar pattern is present with total amplitude $d = 0.060^{+0.010}_{-0.009}$ and it points $\sim$\,125$^{\circ} $ away from the Galatic centre, shown with an asterisk.  This result indicates that the anisotropy has an extragalactic origin.  The growth of the dipole amplitude as a function of energy is shown in Fig.~\ref{f12}.  The data were divided into four bins for $E > 4$ EeV and fitted with the expression $ d = d_{10}(E/10\,\textrm{EeV})^{\beta} $, with $ d_{10} = 0.051 \pm 0.007$ and $\beta = 0.96 \pm 0.16$.  The data are compared to predictions from \cite{Harari2015} for scenarios of extragalactic sources with a density of $10^{-4}$ Mpc$^{-3}$ with a mixed cosmic ray composition which agrees with the ones measured by the Pierre Auger Observatory.  The sources were sampled either from an isotropic distribution or according to the distribution of galaxies in the 2MRS catalog.

\begin{figure}[h]
\begin{minipage}{15pc}
\includegraphics[width=15pc]{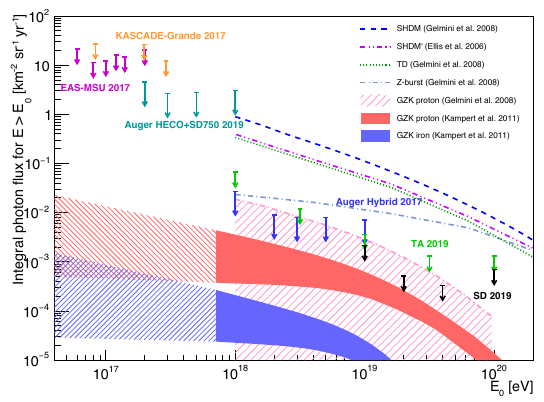}
\caption{ Upper limits on the cosmogenic photon  flux compared to limits from other experiments and to model predictions  \cite{Rautenberg:2019TI}.}
 \labelf{f13} 
\end{minipage}\hspace{1pc}%
\begin{minipage}{15pc}
\includegraphics[width=15pc]{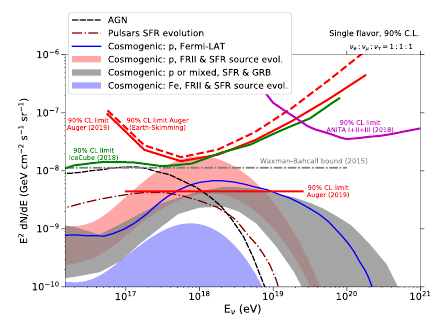}
\caption{ Upper limits on the cosmogenic neutrino  flux compared to limits from other experiments and to model predictions  \cite{Pedreira:2019rM}. }
 \labelf{f14} 
\end{minipage}
\end{figure}

\label{sec:multi}
\section*{Multi-messengers }

The Pierre Auger Observatory is a key detector for multi-messenger physics at EeV energies.  Photons and neutrinos can be discriminated in the background of charged UHECRs with large identification efficiency of about 50\% for photons and more than 85\%  for neutrinos.  Moreover, the continuous monitoring of a large fraction of the sky can be done with good angular resolution.  And finally, the observatory has a unrivalled sensitivity to transient sources if they are located in regions of the sky within  its  field of view. It is relatively easy to identify neutrinos which interact in the lower layers of the atmosphere by looking at the development of the cosmic-ray showers  as a function of the atmospheric depth.  Photon showers develop deeper than cosmic rays in the atmosphere producing less muons making it possible to discriminate them \cite{Aab:2019gra}.  

In Fig.~\ref{f13}   the upper limits on the integral  photon flux from hybrid and SD data are shown together with  results from other experiments.  The search for photons was performed in a energy range of 
three decades due to an exposure of about 40~000~km$^2$~sr~yr and to the analysis of data from the low energy  enhancements SD750 and HEAT.  Eleven candidates were found but since their background hypothesis cannot be excluded with further analysis using proton simulations for the geometry and energy of the candidates, they are conservatively considered as background. 
The green arrows are integral photon upper limits from the SD750 and HEAT extensions assuming a photon flux following $\sim E_{\gamma}^{-2}$  and with no background subtraction.
The limits shown as blue arrows  are from the  hybrid data sample   and  those as  black arrows  are from the SD data sample. Previous data from Auger as well as data from TA, AGASA, Yakutsk, and Haverah Park are included for comparison. The lines and shaded regions give the predictions for top-down models and GZK photon fluxes, respectively. 
 Some top-down scenarios proposed to explain the origin of trans-GZK cosmic rays  are severely constrained.  This analysis was presented in \cite{Abraham:2009qb,Bleve:2016Rt,Aab:2016agp} and updated in \cite{Rautenberg:2019TI}.

  The single flavour 90\% upper limit on the integrated neutrino flux, for an assumed flux $\sim E_{\nu}^{-2}$ and no candidates, is $k_{90} < 4.4 \times 10^{-9}$ GeV cm$^{-2}$ s$^{-1}$ sr$^{-1}$.  It is shown in  Fig.~\ref{f14} as the solid straight red line.  It mostly applies in the energy interval $10^{17}$~eV -- $2.5 \times 10^{19}$~eV for which $\sim 90$\% of the total event rate is expected for the assumed spectral flux.  Also plotted are the upper differential limits. The solid red line is for Auger all channels and flavors. The dashed red line is for the Auger Earth-skimming $\nu_{\tau}$ only.  Similar limits for ANITA and Ice  Cube are displayed along with predictions for several neutrino models.  All limits and fluxes are converted to a single flavor.  This analysis was presented in \cite{Aab:2015kma,Bleve:2016Rt} and updated in \cite{Pedreira:2019rM}.

\label{sec:concl}
\section*{Outlook: AugerPrime}
The Pierre Auger Observatory is being upgraded in order to  extend the composition sensitivity of the experiment  into the flux suppression region.  It will also allow the estimation of the primary mass of the highest energy cosmic rays on a shower-by-shower basis.  The measurement of the mass composition, the search for light primaries at the highest energies, the study of composition-selected anisotropy  and the search for new phenomena including unexpected changes of hadronic interactions are the main objectives of the upgrade \cite{AugerUpgrade2016}. 

Each WCD  will be equipped with a 3.8 m$^2$ slab of plastic scintillator (the Surface Scintillator Detector, SSD) \cite{Taboada:2019+v}, new SD electronics \cite{Nitz:2019i/} and an additional small photomultiplier inside the WCD for the extension of the dynamic range.    In addition, a  radio antenna will be mounted above each WCD for the detection of radio signals from the extensive air showers in the frequency range from 30 to 80 MHz \cite{Pont:2019vo}.  Finally, a UMD will be deployed beside each WCD from the SD750 array to make direct muon measurements possible \cite{Botti:2019zC}. The upgrade is named AugerPrime.  The prototypes of the radio detectors  are now installed in the field and the mass production is underway.
The production and deployment of the AugerPrime detectors and electronics will be completed by 2021, for a data-taking planned up to 2025 and beyond \cite{Castellina:20192A}.

The Pierre Auger Observatory  will continue to provide first quality data that surely will assure its position as one of the major players in the astroparticle physics field.

\bibliographystyle{pepan}
\bibliography{b_CR_other,b_INSPIRE_24Oct2020}

\end{document}